# Computationally Efficient Prediction of Area per Lipid

Vitaly Chaban[1]

MEMPHYS – Center for Biomembrane Physics, Syddansk Universitet, Odense M., 5230, DENMARK

**Abstract**. Area per lipid (APL) is an important property of biological and artificial membranes. Newly constructed bilayers are characterized by their APL and newly elaborated force fields must reproduce APL. Computer simulations of APL are very expensive due to slow conformational dynamics. The simulated dynamics increases exponentially with respect to temperature. APL dependence on temperature is linear over an entire temperature range. I provide numerical evidence that thermal expansion coefficient of a lipid bilayer can be computed at elevated temperatures and extrapolated to the temperature of interest. Thus, sampling times to predict accurate APL are reduced by a factor of ~10.

[1] E-mail: vvchaban@gmail.com

TOC Graphic

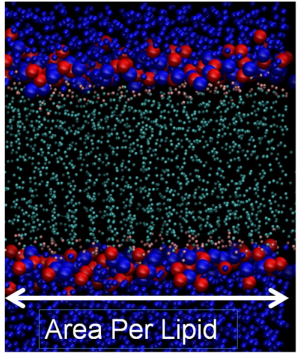

**Research Highlights**

(1) Areas per lipid were computed for pure and mixed lipid bilayers using molecular dynamics.

(2) Area per lipid was predicted using expansion coefficients for lipid bilayers.

(3) Equilibration of bilayer at elevated temperature is much faster than at room temperature.

**Introduction**

Physical chemical properties of self-assembling systems[1-3] and bilayers[4-19] constitute a large body of biophysical research. Computer simulation, in particular molecular dynamics (MD), provides an important contribution to these investigations as an atomistic-precision tool.[8, 18, 20-25] MD simulations of model bilayers have provided comprehensive insights into structure and function of these fascinating systems. In turn, the area per lipid (APL) of a bilayer provides important information about a bilayer or a membrane, because of its high sensitivity to hydrophilic attraction between head groups and hydrophobic repulsion between non-polar hydrocarbon tails. Interaction of head groups with surrounding water or aqueous solution additionally influences APL. Correct prediction of the APL often implies correct prediction of an accurate 2-D density in the bilayer plane, and therefore automatically leads to accurate simulation prediction of other structural properties, such as the lipid tail order parameters, bilayer thickness, electron density profiles, as well as overall phase behavior.

Accurate prediction of an APL via MD simulations is computationally expensive due to slow conformational dynamics of lipids and lipid-like self-assembling systems in the bilayer plane. Normally, hundreds of nanoseconds are required to obtain an equilibrated value even for pure systems, irrespective of the force field employed.[26-28] In terms of wall time, this results in 2-4 weeks long molecular dynamics simulations. Coarse-grained models perform faster, but do not always exhibit the same level of accuracy. Since APL is frequently employed as a descriptor of the bilayer, a method to predict it quickly is necessary. We introduce such a method by using a thermal expansion coefficient which, as it will be demonstrated, is a constant value over an entire fluid phase temperature range of the common bilayers.

Most substances involving solids, liquids, and gases, slightly expand (density decrease) during heating. If an equation of state is available, it can be used to predict the values of the thermal expansion at all the required temperatures and pressures, along with many other state functions. Normally, the equation of state is not available though. Expansion of gases is sensitive to pressure; therefore, it must be determined at constant pressure. In turn, volume change in solids due to external pressure is normally marginal. Pressure dependence can be neglected. Liquids occupy an intermediate niche, which is somewhat closer to the solid case. At a given temperature, a lipid bilayer exists in either liquid or solid phase.[24, 25, 29-33] The solid phase is commonly referred as a gel phase. The characteristic temperature unique for each lipid defines the transition from the gel to the liquid phase. The lipids are constrained to the two dimensional plane of the membrane in both phases, but they diffuse freely in this plane in the liquid phase, whereas they lack kinetic energy to do so in the gel phase. Our current investigation deals with the lipid liquid phase only. That is, all considered temperatures exceed the melting point of the corresponding lipid or mixture.

This work computes the thermal expansion for pure POPC, DPPC, PLPC, POPE bilayers and DPPC/CHOL mixture, which is used to provide a computationally inexpensive, highly precise estimation of APL. The method proposed here will significantly accelerate the modeling and simulation of lipid bilayers and of many complex membrane-associated systems such as interaction and behavior of peripheral and membrane proteins, drugs and other external agents in bilayers. The method should be fully extendable to single phases of other complex condensed matter systems that equilibrate slowly at ambient temperatures.

**Simulation Details**

All APL values reported in the present work have been obtained from molecular dynamics simulations using atomistic (Berger-Tieleman)[34] and coarse-grained (MARTINI)[28] force fields (FFs). The GROMACS program[35] was used to record trajectories. Constant temperature constant pressure ensemble representation with independent lateral and normal weak pressure coupling scheme was used to obtain APL at all temperatures (310-420 K). The initial bilayer structures were obtained from Tieleman's website (http://www.ucalgary.ca/tieleman) and MARTINI website (http://cgmartini.nl/cgmartini). The brute-force calculations of APL were performed using the 400 ns long trajectory at temperatures slightly above the gel-fluid phase transition temperature of the corresponding bilayer. Simulations of APL at the elevated temperatures were performed using the 20 ns long trajectories, of which the first 5 ns were regarded as equilibration (after the temperature increase/decrease by steps of 10 K) and the last 15 ns were used for the APL calculation. Different FF related MD parameters for Berger-Tieleman and MARTINI simulations are briefly detailed below.

Berger-Tieleman FF simulations utilized an integration time-step of 2 fs. Such a time-step was possible due to constraining all covalent bonds involving hydrogen atoms. Electrostatic interactions were computed using direct pairwise Coulomb potential at the separations smaller than 1.0 nm. For all larger distances between the participating atoms, Particle-Mesh-Ewald technique was applied. Note, that the hydrocarbon chains in Berger-Tieleman FF does not use partial point charges, i.e. small fractions of electron charge. The neglected effect is, on the average, compensated by specifically tuned Lennard-Jones (12,6) parameters. The cut-off distance of 1.2 nm for Lennard-Jones potential was employed in conjunction with the shifted force modification between 0.9 and 1.0 nm. The list of neighboring atoms was updated every 20 fs within a radius of 1.0 nm.

MARTINI FF simulations utilized an integration time-step of 30 fs. Berendsen thermostat and barostat[36] have been applied to control temperature and pressure, respectively. The relaxation times are selected in view of computational efficiency and algorithm stability, being 2.0 ps for the thermostat and 4.0 ps for the barostat. The cut-off distance of 1.2 nm for Lennard-Jones potential was employed in conjunction with the shifted force modification between 0.9 and 1.2 nm. The electrostatic interactions were computed using direct pairwise Coulomb potential at the separations smaller than 1.2 nm with a shifted force modification from 0 to 1.2 nm. No electrostatic interactions were considered beyond this cut-off distance, even if the box side length of the system was larger than 2.4 nm. MARTINI FF systematically neglects long-range electrostatic interactions. This solution is implemented for efficiency considerations to the prejudice of physical relevance. MARTINI FF empirical parameters implicitly account for this assumption to provide correct physical properties.

The list of simulated systems and selected results are provided in Table 1. Note, that ready-to-use lipid bilayer configurations were taken from http://www.ucalgary.ca/tieleman. These configurations were previously tested in the works by Tieleman and coworkers at the physiological temperature.[26, 27]

Table 1. The list of simulated systems. Each system was simulated at a variety of temperatures (310-420 K). The system compositions were taken from the corresponding websites, as detailed in the methodology. The original DPPC-CHOL system was enlarged four times in lateral plane to provide better sampling

| System | N(lipid) | N(water) | Force field | Correlation coefficient, $R^2$ | APL, $nm^2$ |
|---|---|---|---|---|---|
| POPC | 128 | 2460 | BERGER-TIELEMAN | 0.9794 | 0.590 (310 K) |
| DPPC | 64 | 3846 | BERGER-TIELEMAN | 0.9789 | 0.567 (320 K) |
| PLPC | 128 | 2453 | BERGER-TIELEMAN | 0.9990 | 0.626 (310 K) |
| POPE | 128 | 6000 | MARTINI | 0.9992 | 0.616 (310 K) |
| DPPC/CHOL | 152+64 | 5600 | MARTINI | 0.9983 | 0.461 (320 K) |

**Results and Discussion**

Area per lipid was simulated at 310-420 K with a temperature increment of 10 K (Figure 1). Linear expansion of lipids with respect to temperature is confirmed. Certain deviations are observed below 350 K. They are due to insufficient sampling (5+15 ns) at these temperatures. Certain deviations are also observed at temperatures above 390 K. They are most likely due to alterations in the properties (density) of water above its normal boiling point. In essence, superheated water is present in these MD systems instead of ordinary liquid water. Such a system is metastable in its nature at ambient pressure. If the points corresponding to lower and higher temperatures are removed, the remaining points can be fitted by a straight line with a high correlation coefficient ($R^2 > 0.97$). The slope of this line is regarded as a thermal expansion coefficient, $k_T$, in nm$^2$ per K. The coefficient is constant over an entire fluid state temperature range. Upon heating, lipid particles begin moving faster and thus maintain a greater spatial separation. The phenomenon of volume increase in response to temperature increase through heat transfer is common in materials. On the contrary, there is a limited group of substances, which contract upon heating within certain temperature range (so-called negative thermal expansion, -$k_T$). For instance, $k_T$ of water drops to zero at ca. 4 K,[37] becoming negative below this temperature. This feature of water has a great number of implications in the living nature. Purified silicon exhibits negative $k_T$ 20 through 120 K.[38] Cubic scandium trifluoride also exhibits such a behavior. This is explained by the quartic oscillation of the fluoride ions.[39] The energy corresponding to the bending strain of the fluoride ion is proportional to the fourth power of the displacement angle. Compare, in the fluorine containing materials with positive $k_T$, this energy is proportional to the square of the displacement. A fluorine atom is bound to two scandium atoms. As temperature increases, the fluorine atom oscillates more perpendicularly to its bonds. Consequently, scandium atoms are drawn together throughout the material resulting in contraction. This property of $ScF_3$ was recorded from 10 K to 1100 K, whereas positive thermal expansion occurs above this temperature.

Cubic scandium trifluoride presents an important example of how anomalous expansion coefficient arises in certain materials. Note, that no material exhibits negative expansion over an entire temperature range. Such a behavior is always confined to certain temperature range, while the same material expands normally in other temperature ranges. Our investigation confirms that lipids belong to a more common group of materials, exhibiting $k_T > 0$ in the fluid phase.

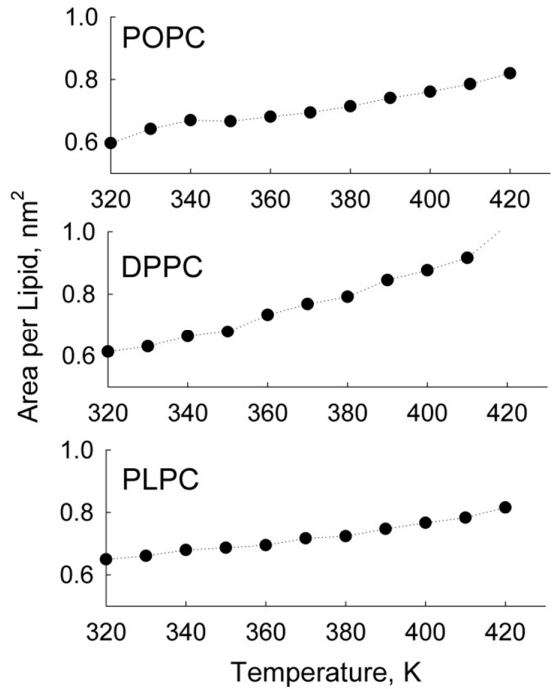

Figure 1. Area per lipid for selected pure lipid bilayers as temperature ranges between 320 and 420 K. Error bars of the depicted values fall within 1% of the absolute value as follows from volume fluctuations in the course of MD. Note, that systematic errors are present for simulations at lower temperatures.

Figure 2 depicts APL fluctuations at the selected temperatures due to thermal fluctuations in the systems during MD simulation. Very fast relaxation of APL is observed in all cases. No drift is observed during 40 ns, suggesting thermodynamic equilibrium. Note, however, that frequency of fluctuations is relatively low. For instance, the average APL at 350 K from 25 to 30 ns is somewhat smaller than the total average. Either longer sampling times (more than 5 ns) must be used or T=350 K must be disregarded in the case of the POPC bilayer. Figure 3 investigates equilibration time of the bilayer upon immediate heating and cooling: from 310 to 380 and vice versa. Upon heating up to 380 K, APL equilibrates within 3 ns. In the case of cooling, APL decreases to 0.48 $nm^2$ during 2 ns, whereas a correct equilibrium value equals to 0.45 $nm^2$. Eventual compression of the mixed lipid bilayer — from 0.48 to 0.45 $nm^2$ per lipid — takes over 100 ns at 310 K (not depicted). This result evidences that the proposed method cannot be substituted by simulated annealing as well as any variation of replica exchange. Indeed, the bilayer can be quickly compressed but achievement of correct APL at 310-320 K is burdened by slow conformational dynamics within this temperature range.

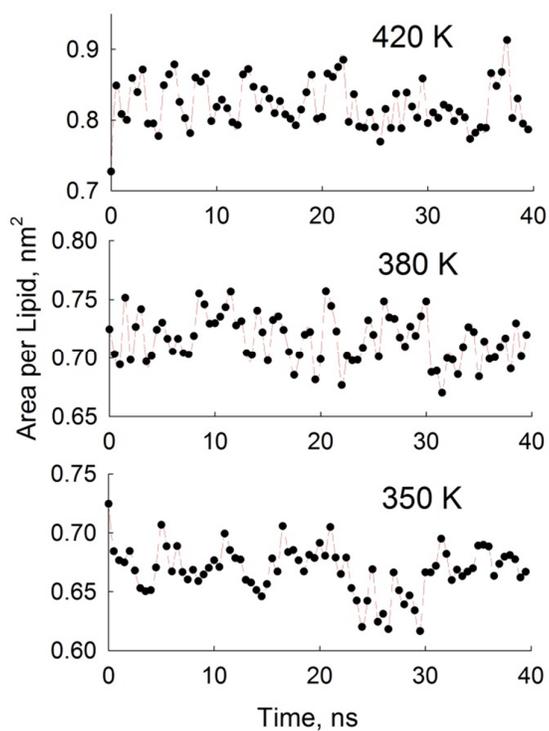

Figure 2. Area per lipid vs. simulated time for POPC bilayer at 420, 380, and 350 K. The plots illustrate fast APL convergence and absence of a longer-term drift.

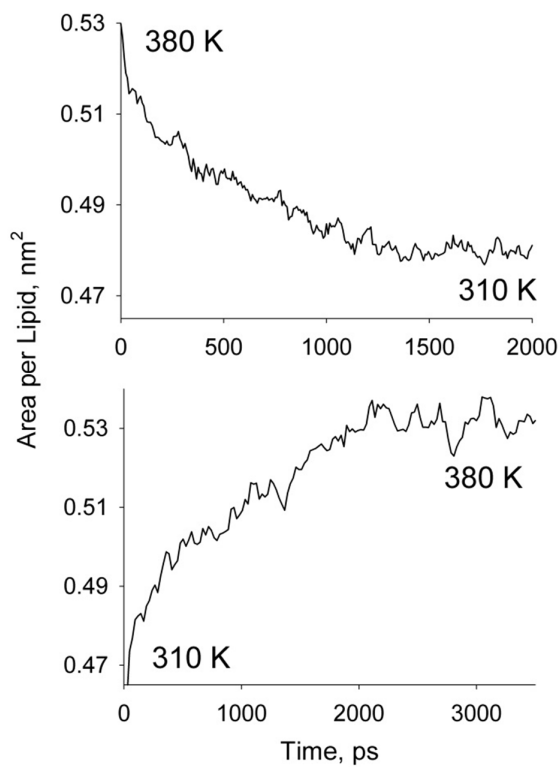

Figure 3. Area per lipid relaxation over time as thermostat reference temperature immediately changes from 380 to 310 K (top) and from 310 to 380 K (bottom). The analysis was done for DPPC/CHOL mixture.

We chose 350, 370, and 390 K to obtain thermal expansion coefficients (Figures 4-5). These points were chosen arbitrarily, as evenly spaced temperatures (dT=20 K), which are high enough to provide APL convergence within 15 ns (Figure 1) and low enough with respect to the critical point of water. It is important to maintain water in the liquid state, as its density and hydrophilic interactions with the head groups are among key factors that define APL. Failure to do so would result in a systematically increased APL. Poorly equilibrated APL values reported in the previous publications are always somewhat larger than the correct ones, but not contrariwise. Table 1 suggests that the points are fitted with a high level of precision evaluated in terms of correlation coefficients ($R^2 > 0.97$). Larger number of APL points at higher temperatures increases $R^2$. Mathematically, APL values derived at two temperatures provide all required information about the derivative, d(APL)/dT. MARTINI FF provides higher $R^2$s, because of simpler potential energy landscape and, therefore, faster accumulation of statistics. The united-atom Berger-Tieleman FF[34] requires longer sampling times and, therefore, performs with a somewhat lesser accuracy — based on 15 ns trajectories. Fully atomistic models would require even longer MD simulations.

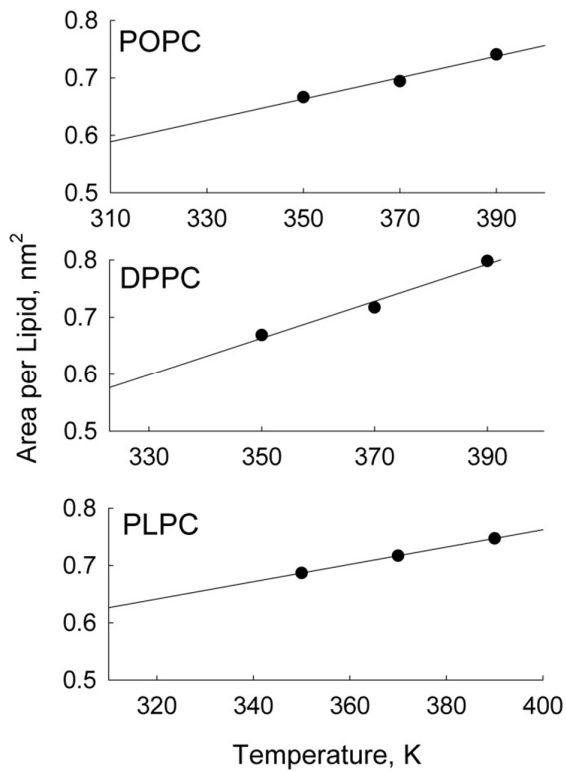

Figure 4. Linear regressions using three APL values at arbitrarily chosen, evenly spaced temperatures. Usage of alternative point for fitting does not bring significant changes to the extrapolated APL.

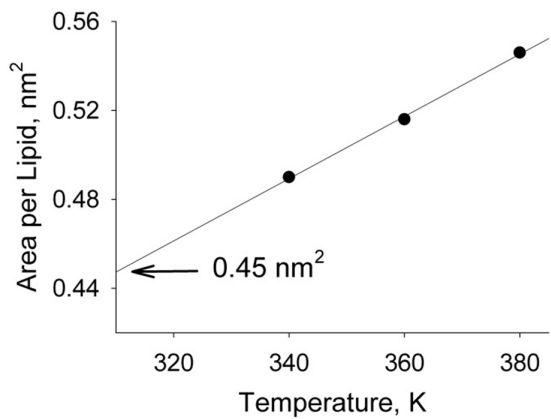

Figure 5. Linear regressions using three APL values of DPPC/CHOL mixture at three evenly spaced temperatures. APL is averaged with respect to all lipid molecules (DPPC+CHOL) in the simulated system.

The derived thermal expansion coefficients amount to $1.9\times10^{-3}$ (POPC), $3.2\times10^{-3}$ (DPPC), $1.5\times10^{-3}$ (PLPC), $1.5\times10^{-3}$ (POPE), $1.4\times10^{-3}$ nm$^2$ K$^{-1}$ (DPPC/CHOL). All coefficients are of the same order of magnitude. These values are in general agreement with experimentally determined $k_T$s for the liquid phase derived from various phospholipids vesicles.[40, 41]

All APL values determined at lower temperatures, obtained using the proposed extrapolation (Figure 6), appear in perfect agreement with the APL values obtained using very long MD simulations at the corresponding temperatures (310-320 K). Recall, we used 400 ns in this work, whereas many works reporting APL from smaller sampling, actually provide incorrect values. In most cases, APL tends to decrease during a long MD run. This is because only during a long MD run lipid molecules find their most energetically favorable conformations. Note, that extrapolation is valid only over the temperature range corresponding to the fluid phase of the bilayer. It can be also applied to predict APLs in supercooled bilayers, if some of them are stable over a significant time scale. The discrepancies are below one percent of the absolute value for all five systems simulated in this work, including a lipid mixture system. These numbers fall within the range of thermal density fluctuations (~1%) using the selected parameters of the thermostat and barostat. Remarkably, the predictions of APL are reliable even in the cases of a small MD system (64 DPPC lipids) and minimum size of the surrounding water phase.

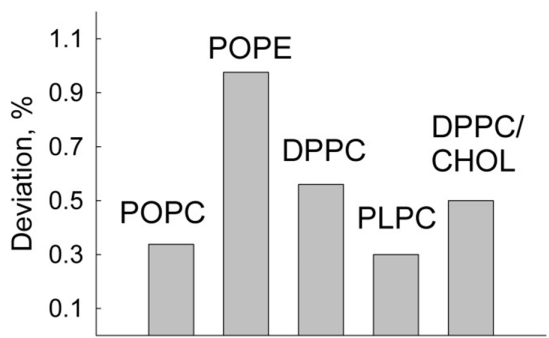

Figure 6. Relative errors of predicted areas per lipid in the five lipid bilayers selected for testing. The relative error is defined as a difference between extrapolated APL and brute-force APL (400 ns sampling) divided by the brute-force APL.

We provide average APL for DPPC/CHOL mixture (Figure 5). This value can be decomposed into components (APL per DPPC lipid and APL per CHOL lipid) under assumptions and approximations. However, such decomposition goes beyond the scope of the present work. The thermal expansion coefficient, $k_T$, for the lipid mixture is a linear combination of $k_T$s of pure lipids, $k_T = k_{T1}x_1 + k_{T2}x_2$, with respect to their fractions, $x_1$, $x_2$. Ideal mixing is assumed in the above equation. However, significant deviations from ideal behavior must be expected in this particular case. Further investigation along these lines is outside our scope, since average APL (Figure 5) provides a sufficient means for our present purpose. Pure CHOL does not engender a bilayer. Nevertheless, its expansion coefficient can be directly obtained from density decrease upon temperature increase, d(density)/dT.

**Conclusions**

The paper demonstrates a computationally efficient method capable of predicting area per lipid for a variety of bilayer systems. The method implementation is straightforward. The proposed method in supported by a clear physical background, which includes (1) thermal expansion of most materials and (2) continuity of properties within a single thermodynamic phase. In turn, sampling speed increases exponentially with temperature increase following Arrhenius-type equations for transport properties, $TP = A \times \exp(-B \times T^{-1})$, where TP is a property, which is proportional to molecular mobility (self-diffusion, heat conductivity), and A, B are empirical constants. We confirm positive expansion coefficients for pure POPC, DPPC, PLPC, POPE, and mixed DPPC/CHOL bilayers. These coefficients are not currently available systematically for common bilayers.

Bilayer simulations emerge nowadays.[4, 20-24, 28, 31, 42, 43] APL is a major bilayer structure property as it is a reliable fingerprint. Therefore, the proposed method will find wide applications in the computational investigations of biophysical systems. Despite clarity and simplicity, it has never been applied to simulate APL of the bilayers. Simulations of mixed bilayers, protein containing bilayers, bilayers encapsulating small molecules inside can be fulfilled. A significant application is a quick screening of the new empirical force fields,[21-23, 28] since reproduction of APL is considered an important reliability test. We achieve an order of magnitude sampling acceleration as compared to the brute-force case. We believe that our report will encourage biophysical community to apply enhanced sampling techniques and, in this way, foster bilayer research.

**Acknowledgments**

The discussed computations have been partially carried at the SDU node of the Danish Center for Scientific Computing. MEMPHYS is the Danish National Center of Excellence for Biomembrane Physics. The Center is supported by the Danish National Research Foundation.

**Author Information**

E-mail for correspondence: vvchaban@gmail.com; chaban@sdu.dk (V.V.C.)